\documentclass[conference]{IEEEtran}
\IEEEoverridecommandlockouts

\usepackage{cite}
\usepackage{amsmath,amssymb,amsfonts}
\usepackage{algorithmic}
\usepackage{textcomp}
\usepackage{xcolor}

\usepackage[T1]{fontenc}
\usepackage{balance}
\usepackage[pdftex]{graphicx}
\graphicspath{{./figures/}}
\DeclareGraphicsExtensions{.pdf,.jpeg,.png}
\usepackage{listings}

\usepackage[absolute]{textpos}
\usepackage[hidelinks]{hyperref}

\begin{document}

\begin{textblock}{15}(0.5,14.9)
{
\noindent\hrulefill

\noindent\fontsize{8pt}{8pt}\selectfont\copyright\ 2018 IEEE. Personal use of this material is permitted. Permission from IEEE must be obtained for all other uses, in any current or future media, including reprinting/republishing this material for advertising or promotional purposes, creating new collective works, for resale or redistribution to servers or lists, or reuse of any copyrighted component of this work in other works. \hspace{5pt} This is the accepted version of: M. Sul\'ir. Integrating Runtime Values with Source Code to Facilitate Program Comprehension. 2018 IEEE International Conference on Software Maintenance and Evolution (ICSME), IEEE, 2018, pp. 743--748. \url{http://doi.org/10.1109/ICSME.2018.00093}

}
\end{textblock}

\title{Integrating Runtime Values with Source Code\\to Facilitate Program Comprehension
\thanks{This work was supported by project KEGA 047TUKE-4/2016 Integrating software processes into the teaching of programming.}
}

\author{\IEEEauthorblockN{Mat\'u\v{s} Sul\'ir}
\IEEEauthorblockA{\textit{Department of Computers and Informatics} \\
\textit{Faculty of Electrical Engineering and Informatics} \\
\textit{Technical University of Ko\v{s}ice}\\
Ko\v{s}ice, Slovakia \\
matus.sulir@tuke.sk}
}

\maketitle

\begin{abstract}
An inherently abstract nature of source code makes programs difficult to understand. In our research, we designed three techniques utilizing concrete values of variables and other expressions during program execution. RuntimeSearch is a debugger extension searching for a given string in all expressions at runtime. DynamiDoc generates documentation sentences containing examples of arguments, return values and state changes. RuntimeSamp augments source code lines in the IDE (integrated development environment) with sample variable values. In this post-doctoral article, we briefly describe these three approaches and related motivational studies, surveys and evaluations. We also reflect on the PhD study, providing advice for current students. Finally, short-term and long-term future work is described.
\end{abstract}

\begin{IEEEkeywords}
integrated development environment, documentation, debugging, dynamic analysis, variables
\end{IEEEkeywords}

\section{Introduction}

In this article, we would like to summarize some of the main results of the thesis \cite{Sulir18integrating}. We also describe the lessons learned and directions for future research.

\subsection{Background}

Maintenance of existing software systems requires the developers to understand the programs of interest. This is accomplished by gradually building a mental model of selected parts of the program \cite{Mayrhauser95program}. One way to build such a mental model is to read the source code lines in the editor. However, the source code provides only a static and abstract view of the program, separated from its runtime properties. To connect these two separate worlds, there exists a large variety of methods, approaches and tools.

In our research, we are particularly interested in three types of activities related to program comprehension.

First, there is a need to find the relevant pieces of code. This process is known as concept location (or feature location \cite{Dit13feature}). Second, to gain an overview of the behavior of the individual methods in the code, developers often read API (Application Programming Interface) documentation \cite{DualaEkoko12asking}. Third, to understand the details of a particular method, the developers can read the source code of the method definition. To alleviate this, many tools try to visually augment the source code directly in the editor to provide additional information in-place \cite{Sulir18visual}.

Dynamic analysis, i.e., the analysis of a running program, is a well-known approach to facilitate software comprehension and maintenance (see, e.g., \cite{Cornelissen09systematic}, \cite{Roethlisberger12exploiting}, \cite{Beck13in}). However, the program execution is usually captured at a high level. The execution is often perceived only as a sequence of method calls, object creations or line executions. For example, none of the feature location approaches described in the articles surveyed by Dit et al. \cite{Dit13feature} analyzed concrete values of local or member variables during executions.

\subsection{Synopsis}

The main goal of our research is to ease program understanding by integrating runtime information with the source code. Particularly, we are focused on concrete values of individual variables and expressions (such as local variables, arguments, return values or member variables). We designed three techniques aiming to help the developers to perform the three aforementioned activities -- searching, documentation reading, and source code reading:

\begin{itemize}
\item RuntimeSearch, a debugger extension which allows for searching a given text in all string expressions in a running program  \cite{Sulir17runtimesearch},
\item DynamiDoc, an automated documentation generator producing sentences with examples of arguments, return values and object state changes collected during executions \cite{Sulir17generating},
\item RuntimeSamp -- an IDE (integrated development environment) plugin showing a sample value for each variable at the end of each line in the source code editor \cite{Sulir18augmenting}.
\end{itemize}

Along with the design of these three tools, we performed supporting empirical studies and conducted related surveys. In the following chapters, we will briefly describe each of the approaches and related findings.

\begin{figure}
\centering
\includegraphics[scale=0.8]{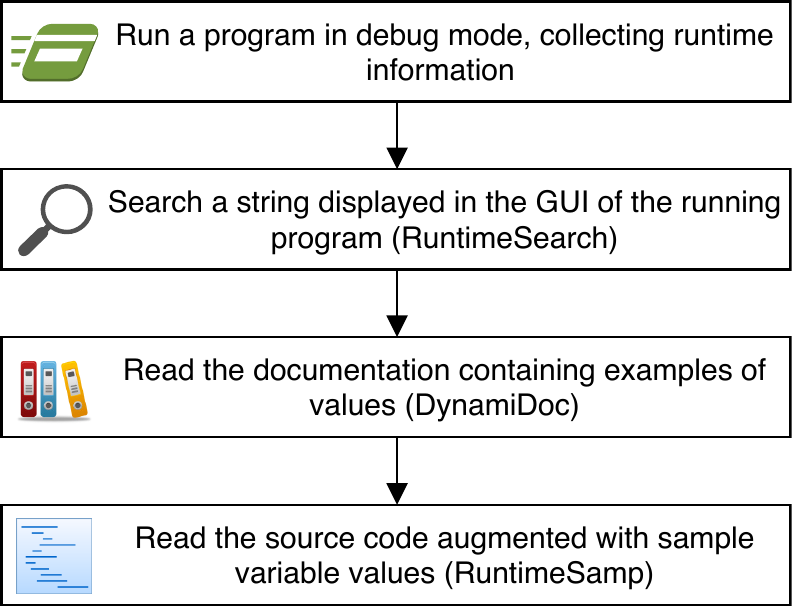}
\caption{An example of a combination of the designed techniques}
\label{f:usage}
\end{figure}

In Fig.~\ref{f:usage}, there is an example of how the three designed techniques might be used together. However, note that each tool is also useful on its own.

\section{Searching}

An important task during software maintenance is to find where the given functionality is implemented. Especially if the software is large, finding an initial investigation point in the codebase is difficult. Although there exists a large number of feature location methods, they are rarely used in practice -- industrial developers prefer traditional approaches such as a textual search in the source code \cite{Damevski16field,Wang11exploratory}.

\subsection{Empirical Study}

The search queries of developers often contain terms obtained by an observation of a running program. For instance, a developer can try to search for a label displayed in the graphical user interface (GUI) of a running application \cite{Roehm15two}. A programmer also tends to ask what part of the code generated the displayed error message \cite{Sillito08asking}.

A naive strategy is to statically search the displayed string in the code as-is. In our small-scale study, we aimed to find to what extent this strategy is sufficient \cite{Sulir16locating}. Four desktop Java applications were scraped to produce a list of strings and words displayed in their GUIs, such as menu items or button labels. We found that about 11\% of strings displayed in the GUIs of running programs were not found in the source code at all, making this strategy ineffective in these cases. More than 24\% of them had more than 100 occurrences, which can be considered too much to be practical for the inspection of all results.

\subsection{RuntimeSearch}

Given this motivation, we designed RuntimeSearch -- a variation of a traditional text search, but for a running program instead of the static source code \cite{Sulir17runtimesearch}. The target application is executed in the debug mode. At any time, the programmer can enter a string into a text field provided by RuntimeSearch. It is subsequently searched in all string-typed expressions being evaluated, such as all string variables and method return values. When a match is found, the program is paused and the traditional IDE debugger is open, offering all standard debugging possibilities, including the inspection of current variable values, stepping and resuming the program. If the current location is irrelevant, we can continue by finding next occurrences.

In contrast to conditional breakpoints, RuntimeSearch searches in all expressions in the program (or the selected packages/classes), not only the selected lines. On the other hand, its capabilities are currently limited. Particularly, it supports only simple string matching. More search options, such as regular expressions, are planned for the future.

\subsection{Evaluation}

First, in a case study on a 350 kLOC (thousands of lines of code) program, we found RuntimeSearch can be useful \cite{Sulir17runtimesearch}:
\begin{itemize}
\item to find an initial point of investigation (e.g. search for a text displayed in the GUI),
\item to search for multiple occurrences of the same string across multiple layers, such as from the GUI through helper methods to file-related routines,
\item to search for non-GUI strings, e.g., texts located in files,
\item to confirm programmer's hypotheses (for instance, trying to find the string ``https://'' if the HTTPS connection is used).
\end{itemize}

The second mentioned point can also be achieved by using a technique we called the ``fabricated text technique'': to enter a dummy text into a part of program accepting textual input (e.g., a text field) and observe the data flow through multiple layers by finding its occurrences using RuntimeSearch.

\begin{figure}
\centering
\includegraphics[scale=0.56]{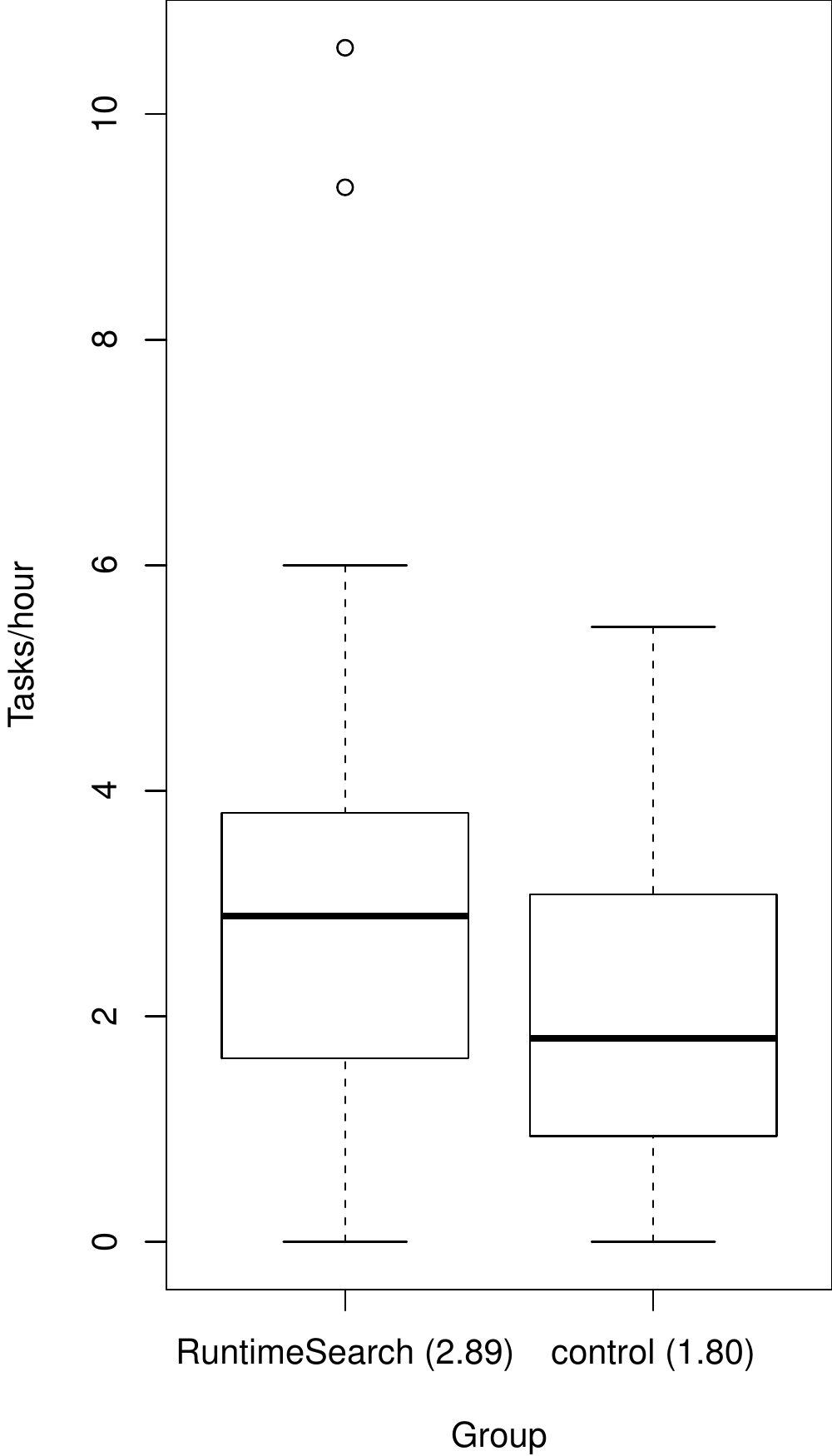}
\caption{Results of the RuntimeSearch controlled experiment}
\label{f:search-efficiency}
\end{figure}

Next, to validate our approach, we performed a (not yet published) controlled experiment with 40 human participants \cite{Sulir18integrating}. One group used RuntimeSearch to perform simple search-focused program maintenance tasks, while the other group could use only standard IDE features. The results of the experiment are in Fig.~\ref{f:search-efficiency}. The treatment group achieved 60\% higher median efficiency in terms of tasks per hour. The difference was statistically significant.

The participants of the experiment were masters students. We received positive feedback from them, multiple students asked whether this tool is publicly available\footnote{Similar to other tools mentioned in this paper, RuntimeSearch is available online: http://sulir.github.io/runtimesearch}. We consider RuntimeSearch a tool which can be soon ready for an industrial transfer -- we plan to publish it on the JetBrains Plugin Repository\footnote{http://plugins.jetbrains.com}. However, first we should make the plugin more production-ready: clean the code, reduce manual steps required to setup the plugin, write the documentation, etc.

\section{Documentation}

While API documentation is a useful resource for programmers, writing it and keeping it consistent with the source code requires a huge effort. Therefore, many automated approaches to documentation generation were devised. However, they traditionally process only the static source code or artifacts like mailing lists \cite{Nazar16summarizing}. Although there exist documentation generators utilizing runtime information, they are specialized -- e.g., FailureDoc \cite{Zhang11automated} for failing unit tests or SpyREST \cite{Sohan15spyrest} for RESTful (Representational State Transfer) APIs.

\subsection{DynamiDoc}

We designed DynamiDoc, an example-based documentation generator utilizing runtime information collected during unit test executions or debugging \cite{Sulir17generating}. For each method (function), it collects:
\begin{itemize}
\item string representations of arguments and return values,
\item the string representations of the target object ({\small\texttt{this}}) before and after calling the given method,
\item and thrown exception types.
\end{itemize}

The representations of objects are obtained using the standard {\small\texttt{toString()}} method in Java, which has an alternative in almost all languages.
 
Then, using a decision table with sentence templates, DynamiDoc generates documentation sentences containing examples of these values. For instance, an excerpt from the documentation of the method Range.lowerBoundType() from Google Guava\footnote{https://github.com/google/guava} may look like this:

{\small\begin{lstlisting}[basicstyle=\ttfamily,mathescape]
When called on (5..8), the method
 returned OPEN.
When called on [5..8), the method returned
 CLOSED.
\end{lstlisting}}

\subsection{Evaluation}

Using a qualitative evaluation \cite{Sulir17generating}, we found out DynamiDoc is  particularly useful for the documentation of utility methods and data structures. On the other hand, methods which manipulate classes not having the {\small\texttt{toString()}} method meaningfully overridden and methods interacting with the external world are not the best candidates for DynamiDoc documentation.

We also performed a preliminary quantitative evaluation \cite{Sulir17source}. We found that on average, one documentation sentence has 10\% of the length of the method it describes, so it is sufficiently succinct. By manually inspecting a sample of documentation sentences, we found 88\% of the described objects have the {\small\texttt{toString()}} method overridden. Therefore, we fulfilled basic prerequisites for the usefulness of this approach.

\section{Augmentation}

Since an understanding of a program only by reading its source code is difficult, many tools augment it with various metadata -- from manually written notes through performance data to information about related emails.

\subsection{Surveys}

In our article \cite{Sulir17labeling}, we described a taxonomy of source code labeling. The taxonomy consists of four dimensions: source (where the metadata come from, such as static or dynamic analysis), target (granularity -- whole method, line, etc.), presentation (in the editor or a separate tool) and persistence.

Then we performed a systematic mapping study \cite{Sulir18visual}, summarizing existing tools which visually augment the textual source code editor with various icons, graphics and textual labels. We found more than 20 tools augmenting the code with runtime information, but very few of them aim to display examples of concrete variable values. IDE sparklines \cite{Beck13visual} are limited to numeric variables, Debugger Canvas \cite{DeLine12debugger} requires the developer to manually select individual states during debugging and the prototype by Kr\"amer et al. \cite{Kraemer14how} suffers from scalability issues. Tralfamadore \cite{Lefebvre09tralfamadore,Bradley10ide} displays only arguments and return values.

\subsection{RuntimeSamp}
\label{s:runtimesamp}

\begin{figure*}
\centering
\includegraphics[scale=0.45]{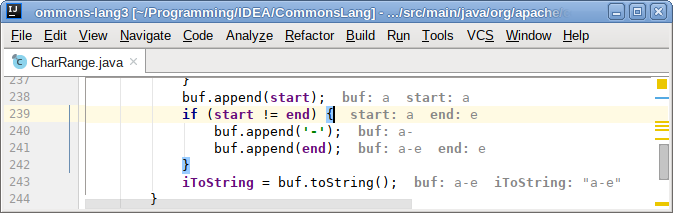}
\caption{RuntimeSamp showing source code augmented with sample variable values (gray)}
\label{f:runtimesamp}
\end{figure*}

Our IDE extension RuntimeSamp \cite{Sulir18augmenting} collects a few sample values of each variable during normal executions of a program by a developer, such as testing or debugging. Then, at the end of each line, one sample value is shown for each variable read or written on the given line. A demonstration, showing an excerpt from the Apache Commons Lang\footnote{https://commons.apache.org/lang/} library can be seen in Fig.~\ref{f:runtimesamp}. The idea behind the tool is that concrete variables should help the developers to get the ``feeling'' of runtime and concreteness in the inherently abstract and static source code.

Compared to DynamiDoc, RuntimeSamp provides more fine-grained data -- it displays information for individual lines and variables instead of whole methods. Furthermore, it is an interactive IDE extension, while DynamiDoc generates static textual documentation.

In our article \cite{Sulir18augmenting}, we asked 7 questions which should be answered for RuntimeSamp to be useful in practice:

\begin{itemize}
\item How to represent complicated objects succinctly?
\item When should we capture the variable values (e.g., is one value per line sufficient)?
\item If one line is executed more than once, how to decide which iteration to display?
\item How to detect and present such iterations?
\item How to keep the time overhead reasonable during the data collection?
\item Is it necessary to filter the displayed variables?
\item When to invalidate the data?
\end{itemize}

For now, we answered these questions mainly in naive ways. To display the values of objects, we use their standard string representations ({\small\texttt{toString}}). We capture the values at the end of each line. Since we consider the caret (text cursor) as an implicit pointer to the programmer's focus point, the first iteration which covers the line at the cursor is always displayed. An iteration is defined as a forward execution (without backward jumps) within one method. When collecting one sample value for each variable, the time overhead is about 78--213\%, which is not prohibitive, but certainly requires an improvement. The measurement was performed using the DaCapo benchmark \cite{Blackburn06dacapo}. We filter the displayed data using a simple rule to prevent redundancy and invalidate all data on any edit (which is only a preliminary solution).

\section{Lessons Learned}

In this section, we would like to describe reflections on the PhD study and advice for other students.

\subsection{Seek Collaboration}

Some of the most valuable publications (e.g., \cite{Sulir16recording}) during the PhD study were written in collaboration with other members of our research group. More people can afford to complete more time-consuming tasks -- this is particularly true if they can be easily divided to sub-tasks, such as certain kinds of controlled experiments or systematic reviews.

Since international collaboration is not an integral part of the research process at our institution, and we did not actively seek such a collaboration, none of the papers included in the dissertation was co-authored by people outside our research group. Therefore, cooperation with other institutions is planned in the near future. A good piece of advice for students is to actively search for opportunities to collaborate with people with similar research ideas during their studies, e.g., at conferences.

\subsection{Focus on Your Topic}

Although collaboration is useful, it can be also considered a double-edged sword. Since the persons you collaborate with may have slightly different research interests than you, the cooperation with them can act as a distraction from the main goals of your thesis. This may make the process of your dissertation completion challenging: You will be left with an option to either make your dissertation topic too broad or exclude a large number of valuable papers from the thesis.

Of course, collaboration is not the sole cause of distraction from the thesis topic. During the initial periods of the PhD study, we had multiple potential ideas for the dissertation topic and we even tried to pursue some of them although they had little in common. While this resulted in some interesting research results (e.g., about build system failures \cite{Sulir16quantitative}), it also delayed the progress on the main topic.

\section{Future Work}

Finally, we will present our short-term future research tasks and long-term visions.

\subsection{Short-Term Goals}

Currently, we are working on the first question mentioned in section~\ref{s:runtimesamp}: How to represent an object, consisting of many properties, on a limited space?

Before considering graphical representations, let us focus on the textual ones. The solution used in RuntimeSamp (and also in DynamiDoc) is to convert it to a string using a standard ``toString''-like method, available in many languages, including Java. However, this representation must be written manually by the programmers, which is a reason it is sometimes left with its default (useless) implementation. Using machine learning, we try to automatically generate the string representation of objects, listing only subsets of their member variables which are considered important by programmers.

Another short-term goal is to evaluate DynamiDoc and RuntimeSamp using experiments with human participants.

\subsection{Long-Term Goals}

The first long-term goal is to extend the object representation question to graphical representations. We can recognize two extremes: On one end, there are generic tree-based and graph-based (such as DDD \cite{Zeller96ddd}) visualizations displaying all properties of the objects, suitable for any kind of data, but revealing little domain-specific information. On the other hand, approaches such as Moldable Inspector allow the developers to craft graphical representations perfectly suitable for a particular domain, but they require manual coding effort \cite{Chis15moldable}. Finding a right compromise between these two extremes is the challenge we would like to address next. This can be even more complicated if we consider not only one state, but also a difference of two or multiple states.

Our main long-term goal is to blend the activities of source code reading/editing and an observation of the runtime properties of the application, so that the line between them will be almost indistinguishable.

One of the research areas aiming to clear this boundary is the area of live programming systems. A large amount of work was done in this field -- from the design of live programming languages \cite{McDirmid07living,Sorensen10programming} and their visual augmentation \cite{Swift13visual} to experiments \cite{Kraemer14how} and integration with unit testing \cite{Imai15making}, just to name a few advances.

Although live-coding ideas are innovative and exciting, a majority of the approaches are looking at live programming from the ``clean slate'' perspective: They do not try to integrate live features into existing mainstream programming languages and IDEs. Even when they do, the ideas are often presented on ``toy examples'', with their application on large industrial systems being disputable.

Note that in reality, it is impractical to throw away existing systems, libraries, the knowledge of programmers and begin from scratch. Therefore, our vision is to gradually improve the experience of developers regarding the connection of source code and runtime in the existing languages and IDEs, without disrupting their current workflow.

We consider RuntimeSamp to be the first step toward our ambitious goal. After improving the object representation, we would like to focus on the data invalidation problem. Instead of deleting all data dependent on the changed parts, we would like to recompute them whenever possible. To prevent cognitive overload, showing only task-relevant runtime information will be necessary. Finally, sufficient performance improvements could make the approach suitable for industrial use.

\bibliography{icsme}
\balance

\end{document}